# Self Excitation of the Tunneling Scalar Field in False Vacuum Decay

Takashi Hamazaki[1], Misao Sasaki[2]
Takahiro Tanaka[3*] and Kazuhiro Yamamoto[1*]

[1] *Department of Physics, Kyoto University*
*Kyoto 606-01, Japan*
[2] *Department of Earth and Space Science,*
*Osaka University, Toyonaka 560, Japan*

and
[3] *Uji Research Center, Yukawa Institute for Theoretical Physics,*
*Kyoto University, Uji 611, Japan*

A method to determine the quantum state of a scalar field after $O(4)$-symmetric bubble nucleation has been developed recently. The method has an advantage that it concisely gives us a clear picture of the resultant quantum state. In particular, one may interpret the excitations as a particle creation phenomenon just as in the case of particle creation in curved spacetime. As an application, we investigate in detail the spectrum of quantum excitations of the tunneling field when it undergoes false vacuum decay. We consider a tunneling potential which is piece-wise quadratic, hence is simple enough to allow us an analytical treatment. We find a strong dependence of the excitation spectrum upon the shape of the potential on the true vacuum side. We then discuss features of the excitation spectrum common to general tunneling potentials not restricted to our simple model.

## I. INTRODUCTION

There is a growing interest in the quantum state of a scalar field inside a vacuum bubble that appears after false vacuum decay. This is because the spacetime region inside the bubble may be considered as a homogeneous and isotropic open universe and, when combined with a class of inflationary models, there is a possibility that our universe is entirely contained in a single bubble and the present large scale structure of the universe with a low density parameter, $\Omega_0 \sim 0.1$, may be explained [1–6]. In the standard scenario, inflation solves both the horizon and flatness problems simultaneously. But in the new one-bubble scenario, these problems are solved by two different inflationary stages of the universe. In the first inflationary stage, the universe is in a false vacuum. If this stage lasts long enough, the universe may be approximated by a de Sitter spacetime with high accuracy when the false vacuum decay occurs through quantum tunneling. This implies the decay is mediated by the $O(4)$-symmetric Euclidean bounce, which consequently ensures that the spacetime after the decay has $O(3,1)$-symmetry [7]. In other words, the region inside a nucleated bubble is a homegeneous and isotropic open universe. Then if the vacuum energy inside the bubble is nonvanishing, the second inflationary stage follows and a large amount of entropy will be produced at the end of the second inflation to solve the flatness problem.[†]

---

*JSPS junior fellow
[†]Here by the flatness of the universe we mean $\Omega_0$ is not greatly different from unity.

In order for this one-bubble inflation scenario to be successful, it is then a matter of great importance that if the quantum fluctuations of a scalar field inside a nucleated bubble can account for the observed large scale structures of the universe as well as for the detected anisotropies of the cosmic microwave background. As has been calculated in Refs. [4,5], the information of the quantum state at the beginning of the second inflation inside a nucleated bubble remains until today and is reflected in the CMB anisotropies on large angular scales, particularly in those corresponding to supercurvature scales at matter-radiation decoupling. It is thus of particular interest to clarify properties of the quantum state inside a nucleated bubble and how they are brought forth through false vacuum decay.

In one of our previous papers [15], we have investigated this problem by assuming that the quantum state is in the Euclidean vacuum (Bunch-Davies vacuum). This means we have neglected the effect of quantum excitations through tunneling process but only taken into account the effect of the background spacetime curvature. Although one can construct a model in which this is a good approximation, such as the one advocated recently by Linde [6], it will not be so if there is only one scalar field and it is responsible both for the false vacuum decay and for the second inflation.

On the other hand, efforts to understand the quantum excitations during tunneling process was first made by Rubakov [8] and then by Vachaspati and Vilenkin [9]. Subsequently, based on the multidimensional wave function approach [10,11,14], we have investigated quantum excitations produced through false vacuum decay [12,13]. In particular, in [13], we have presented a systematic method to evaluate the degree of quantum excitations and calculated the excitation spectrum for simple models. However, there we have made several assumptions which may not be relevant for the one-bubble inflation scenario. One of them is the neglection of the background spacetime curvature. Another is that we did not consider the excitations of the tunneling scalar field itself but some other one that couples to the tunneling field through the mass term. Although these simplifications made it easy to understand gross features of the quantum state inside the bubble, we have to admit that it is far from complete. For example, if one considers the excitations of the tunneling field, there exists inevitably a region of negative mass-squared around the top of the potential barrier and that may affect the results seriously.

In this paper, as a step toward full understanding of the matter, we extend our previous analysis [13] and investigate the quantum excitations of the tunneling field itself through false vacuum decay. However, for simplicity, we neglect the background curvature and assumes the Minkowski background. The paper is organized as follows. In section 2, to make the paper self-contained, we briefly review our formalism for determining the quantum state after false vacuum decay. Our formalism uses the fact that the excitations can be concisely described in the language of particle creation due to a varying mass, as that in curved spacetime. In section 3, we present a model of the tunneling potential which is piece-wise quadratic, hence is simple enough to allow us an analytical treatment but is expected to retain the essential feature specific to the excitations of a tunneling field. Then we express the resulting quantum state in the language of particle creation and give a complete (but complicated) formula of the particle spectrum for this model. In section 4, based on the formulae derived in section 3, we evaluate the particle spectrum for various model parameters in detail. For limiting values of the parameters, we give analytical expressions for the particle spectrum. For other values of the parameters we show the results obtained numerically. In section 5, we discuss possible role of discrete modes which are associated with the vibration of the bubble wall. Although these modes cannot be interpreted as particle modes, we argue that they will also contribute to the quantum state inside the bubble but those with spherical harmonic indices $\ell = 0$ and 1, which represent translational degrees of freedom of the bubble location and gives rise to divergence in the two-point function, will be absorbed into the metric perturbation when gravitational degrees of freedom are taken into account. Finally, section 6 is devoted to summary and future issues.

Throughout this paper, the metric signature for Lorentzian spacetime is $(-+++)$ and the units $\hbar = 1$ and $c = 1$ are used.

## II. REVIEW OF FORMALISM

To begin with, we give a brief sketch of our method for solving the evolution of the quantum state through false vacuum decay. We use the multidimensional WKB wave function approach originally developed by Banks, Bender and Wu [10] and de Vega, Gervais and Sakita [11], and recently elaborated by us [14]. We consider a scalar field $\phi$ in the Minkowski spacetime whose action is given by

$$S = \int d^4\mathbf{x} \left[ -\frac{1}{2} \partial_\mu \phi \, \partial^\mu \phi - V(\phi) \right] \quad , \tag{2.1}$$

where $V(\phi)$ is a tilted double-well type potential as shown in Fig. 2. We consider the situation in which the field is initially at the false vacuum ($\phi = \phi_F$) and decays toward the true vacuum ($\phi = \phi_T$) by quantum tunneling.

In order to trace the evolution of the quantum state through false vacuum decay, we construct a quasi-ground state wave functional which is an energy eigenstate of the time-independent Schrödinger equation and is exponentially close

FIG. 1. A schematic picture of the bounce solution.

In the next WKB order, quantum fluctuations around the bounce solution come into play. Setting $\phi = \phi_B + \varphi$, the WKB wave functional to this order is given by [11,14]

$$\Psi[\phi(\cdot)] = \mathcal{N}(T) \exp\left[-\int^T dT' \int d^3\mathbf{x} \left[\frac{1}{2}\left(\frac{\partial \phi_B}{\partial T'}\right)^2 + \frac{1}{2}\left(\frac{\partial \phi_B}{\partial \mathbf{x}}\right)^2 + V(\phi_B)\right]\right]$$
$$\times \exp\left[-\frac{1}{2} \int\int d^3\mathbf{x} d^3\mathbf{y} \varphi(\mathbf{x}) \Omega(\mathbf{x},\mathbf{y};T) \varphi(\mathbf{y})\right] \quad, \tag{2.5}$$

where

$$\Omega(\mathbf{x},\mathbf{y};T) := \int d^3\mathbf{k} \frac{\partial g_\mathbf{k}(\mathbf{x},T)}{\partial T} g_\mathbf{k}^{-1}(\mathbf{y},T) \quad. \tag{2.6}$$

The function $g_\mathbf{k}(\mathbf{x},T)$, which we call the mode function, satisfies the field equation for $\varphi$ on the background $\phi_B$,

$$\left[\frac{\partial^2}{\partial T^2} + \sum_{i=1}^3 \frac{\partial^2}{\partial x^{i2}} - V''(\phi_B(T^2 + \mathbf{x}^2))\right] g_\mathbf{k}(\mathbf{x},T) = 0 \quad, \tag{2.7}$$

and the inverse $g_\mathbf{k}^{-1}(\mathbf{x},T)$ is defined by

$$\int d^3\mathbf{x} g_\mathbf{k}(\mathbf{x},T) g_{\mathbf{k}'}^{-1}(\mathbf{x},T) = \delta^3(\mathbf{k}-\mathbf{k}') \quad. \tag{2.8}$$

To solve Eq. (2.7), we need to set an appropriate boundary condition. At false vacuum $\phi_B = \phi_F$ ($T = \tau \to -\infty$), the second line of the wave functional (2.5) must coincide with that of the would-be ground state at false vacuum. This requirement determines the boundary condition at $\tau \to -\infty$ as

$$g_\mathbf{k}(\mathbf{x},\tau) \to e^{-i\mathbf{k}\mathbf{x} + \omega_\mathbf{k}\tau} \quad, \tag{2.9}$$

where $\omega_\mathbf{k} := \sqrt{\mathbf{k}^2 + V''(\phi_F)}$. In the Lorentzian region $\mathcal{M}$, the function $g_\mathbf{k}(\mathbf{x},it)$ is given by the analytic continuation of $g_\mathbf{k}(\mathbf{x},\tau)$ through $\tau = 0$. After all, the problem of constructing the WKB wave functional $\Psi$ reduces to the problem of solving Eq.(2.4) and Eq.(2.7).

In $\mathcal{M}$, the state of $\varphi$ described by the second line of Eq.(2.5) has a simple interpretation. In the second quantization picture (regarding $t$ as real physical time), $\varphi$ is represented as a field operator,

$$\varphi(\mathbf{x},t) = \int d^3\mathbf{k} \left(u_\mathbf{k}(\mathbf{x},t) b(\mathbf{k}) + \overline{u_\mathbf{k}(\mathbf{x},t)} b^\dagger(\mathbf{k})\right) \quad, \tag{2.10}$$

where $b_\mathbf{k}$ and $b_\mathbf{k}^\dagger$ are the annihilation and creation operators, respectively, relative to a state $|\tilde{0}\rangle$ annihilated by $b_\mathbf{k}$, and a bar denotes complex conjugation.. The mode functions $u_\mathbf{k}(\mathbf{x},t)$ satisfy the field equation and are normalized with respect to the Klein-Gordon inner product, but are not necessarily positive frequency functions. Hence the 'vacuum' $|\tilde{0}\rangle$ annihilated by $b(\mathbf{k})$ is not generally an eigenstate of the Hamiltonian unlike the Minkowski vacuum $|0\rangle$. The former is a squeezed state over the latter, given by a Bogoliubov transformation, and contains non-vanishing spectrum of excited particles. It is known that the wave functional for the state $|\tilde{0}\rangle$ is given by

$$\langle \varphi(\cdot)|\tilde{0}\rangle \propto \exp\left[-\frac{1}{2} \int\int d^3\mathbf{x} d^3\mathbf{y} \varphi(\mathbf{x}) \Omega(\mathbf{x},\mathbf{y};t) \varphi(\mathbf{y})\right] \tag{2.11}$$

where

$$\Omega(\mathbf{x},\mathbf{y};t) := \int d^3\mathbf{k} \frac{1}{i} \frac{\partial \overline{u_\mathbf{k}(\mathbf{x},t)}}{\partial t} \overline{u_\mathbf{k}^{-1}(\mathbf{y},t)} \quad. \tag{2.12}$$

This expression is nothing but the second line of Eq.(2.5) if $\overline{u_\mathbf{k}(\mathbf{x},t)}$ is identified with $g_\mathbf{k}(\mathbf{x},it)$. Since $\Omega(\mathbf{x},\mathbf{y};t)$ is invariant under linear transformations of $g_\mathbf{k}(\mathbf{x},it)$, we then conclude $\overline{u_\mathbf{k}(\mathbf{x},t)} = \sum_{\mathbf{k}'} c_{\mathbf{k}\mathbf{k}'} g_{\mathbf{k}'}(\mathbf{x},it)$ for some $c_{\mathbf{k}\mathbf{k}'}$ with $\det c_{\mathbf{k}\mathbf{k}'} \neq 0$. Thus the quantum state of $\varphi$ after tunneling is a squeezed state over the Minkowski vacuum determined by the mode function $u_\mathbf{k}(\mathbf{x},t)$.

In order to find the mode function $u_{\mathbf{k}}(\mathbf{x},t)$, it is convenient to use the coordinates which respect the symmetry of the background $\phi_B$, i.e., $O(4)$ in $\mathcal{E}$ and $O(3,1)$ in $\mathcal{M}$ [13]. In the Euclidean region $\mathcal{E}$, we use $(\xi_E, \chi_E, \theta, \varphi)$ where $(\theta, \varphi)$ are the usual two-dimensional spherical coordinates, and $(\xi_E, \chi_E)$ are related to $r = |\mathbf{x}|$ and $\tau$ as

$$r = \xi_E \sin \chi_E, \quad \tau = -\xi_E \cos \chi_E,$$
$$0 \leq \chi_E \leq \frac{\pi}{2}, \quad 0 \leq \xi_E < \infty \quad . \tag{2.13}$$

Then the Euclidean metric takes the form,

$$ds_E^2 = d\xi_E^2 + \xi_E^2(d\chi_E^2 + \sin^2 \chi_E d\Omega^2). \tag{2.14}$$

In the Lorentzian region $\mathcal{M}$, we have two distinctive spacetime regions characterized by the action of $O(3,1)$. They are separated by the light cone that expands from $r = t = 0$. The spacelike region is called the (spherical) Rindler space, and the (future) timelike region the Milne universe. Since we are interested in the quantum state inside the bubble, we focus on the region corresponding to the Milne universe. Then the coordinates in the Milne universe are obtained by the replacement $(\xi_E, \chi_E) \to (-i\xi, i\chi)$, which yields

$$r = \xi \sinh \chi, \quad t = \xi \cosh \chi,$$
$$0 < \xi < \infty, \quad 0 < \chi < \infty \quad . \tag{2.15}$$

The metric of the Milne universe is given by

$$ds_M^2 = -d\xi^2 + \xi^2(d\chi^2 + \sinh^2 \chi d\Omega^2). \tag{2.16}$$

It is to be reminded that $\phi_B$ is constant on the $\xi_E = const.$ (or $\xi = const.$) hypersurface. With these coordinates the Laplacian (or d'alambertian) which appears in the field equation is rewritten as

$$\frac{\partial^2}{\partial \tau^2} + \sum_{i=1}^{3} \frac{\partial^2}{\partial x^{i2}} = \frac{\partial^2}{\partial \xi_E^2} + \frac{3}{\xi_E}\frac{\partial}{\partial \xi_E} + \frac{1}{\xi_E^2}\mathbf{L}_E^2 \quad ,$$
$$\mathbf{L}_E^2 := \frac{1}{\sin^2 \chi_E}\frac{\partial}{\partial \chi_E}\left(\sin^2 \chi_E \frac{\partial}{\partial \chi_E}\right) + \frac{1}{\sin^2 \chi_E}\mathbf{L}_\Omega^2 \quad , \tag{2.17}$$

in the Euclidean region, and as

$$-\frac{\partial^2}{\partial t^2} + \sum_{i=1}^{3} \frac{\partial^2}{\partial x^{i2}} = -\frac{\partial^2}{\partial \xi^2} - \frac{3}{\xi}\frac{\partial}{\partial \xi} + \frac{1}{\xi^2}\mathbf{L}^2 \quad ,$$
$$\mathbf{L}^2 := \frac{1}{\sinh^2 \chi}\frac{\partial}{\partial \chi}\left(\sinh^2 \chi \frac{\partial}{\partial \chi}\right) + \frac{1}{\sinh^2 \chi}\mathbf{L}_\Omega^2 \quad , \tag{2.18}$$

in the Lorentzian region, where $\mathbf{L}_\Omega^2$ is the Laplacian on the unit two sphere.

As the $O(4)$-symmetric bounce solution is considered, Eq.(2.4) reduces to

$$\left[\frac{d^2}{d\xi_E^2} + \frac{3}{\xi_E}\frac{d}{d\xi_E}\right]\phi_B(\xi_E) = V'(\phi_B(\xi_E)) \quad , \tag{2.19}$$

in $\mathcal{E}$. The boundary condition is

$$\phi_B(\xi_E) \to \phi_F \quad \text{for} \quad \xi_E \to \infty,$$
$$\frac{d\phi_B}{d\xi_E}(0) = 0. \tag{2.20}$$

In $\mathcal{M}$, the equation for $\phi_B$ is obtained by replacing $\xi_E$ by $-i\xi$ in Eq.(2.19), which describes the evolution of $\phi_B$ inside a nucleated bubble.

Given the background solution $\phi_B$, we need to solve for $g_{\mathbf{k}}$. For this purpose, we expand $g_{\mathbf{k}}(\mathbf{x}, it)$ in terms of harmonic functions on the three dimensional unit hyperboloid,

$$Y_{p\ell m}(\chi,\Omega) = f_{p\ell}(\chi)Y_{\ell m}(\Omega)\,; \quad f_{p\ell}(\chi) = \left|\frac{\Gamma(ip+\ell+1)}{\Gamma(ip)}\right|\frac{1}{\sqrt{\sinh\chi}}P_{ip-1/2}^{-\ell-1/2}(\cosh\chi), \tag{2.21}$$

which is an eigenfunction of the Laplacian operator $-\mathbf{L}^2$ with eigenvalue $1+p^2$,

$$-\mathbf{L}^2 Y_{p\ell m} = (1+p^2)Y_{p\ell m}\,. \tag{2.22}$$

Then $g_{\mathbf{k}}(\mathbf{x}, it)$ is expanded as

$$g_{\mathbf{k}}(\mathbf{x},it) = \int_0^\infty dp \sum_{\ell m} \lambda(\mathbf{k};p\ell m)\overline{u_{p\ell m}(\xi,\chi,\Omega)}; \quad \overline{u_{p\ell m}(\xi,\chi,\Omega)} = \frac{\overline{F_p(\xi)}}{\xi}\overline{Y_{p\ell m}(\chi,\Omega)}, \tag{2.23}$$

where we have associated the complex conjugate of $u_{p\ell m}$ with $g_{\mathbf{k}}$ in accordance with the general discussion. We mention that $p$ corresponds to a comoving wavenumber in the Milne universe and $p=1$ to the spatial curvature scale on the $\xi = const.$ hypersurface. We then consider the analytic continuation of $\overline{u_{p\ell m}}$ to the Euclidean region by $\xi \to i\xi_E$ ($0 \le \xi_E$) and $\chi \to -i\chi_E$ ($0 \le \chi_E \le \pi/2$). In $\mathcal{E}$, Eq.(2.7) for $g_{\mathbf{k}}$ reduces to the equation for $G_p(\xi_E) := F_p(i\xi_E)$,

$$\left[\frac{\partial^2}{\partial\xi_E^2} + \frac{1}{\xi_E}\frac{\partial}{\partial\xi_E} + \frac{p^2}{\xi_E^2} - V''(\phi_B(\xi_E))\right]G_p(\xi_E) = 0\,. \tag{2.24}$$

It has been shown that the boundary condition (2.9) for $g_{\mathbf{k}}(\mathbf{x},\tau)$ at $\tau \to -\infty$ means the regularity of $g_{\mathbf{k}}$ on the $\tau < 0$ half of the complex $T$-plane. Since $f_{p\ell}(-i\chi_E)$ is regular for $0 \le \chi_E \le \pi/2$, this condition is translated to the boundary condition for $G_p(\xi_E)$ that [13]

$$G_p(\xi_E) \to K_{ip}(m\xi_E) \text{ for } \xi_E \to \infty. \tag{2.25}$$

In $\mathcal{M}$, $F_p(\xi)$ is given by solving Eq.(2.24) with $\xi_E$ replaced by $-i\xi$ and with the asymptotic boundary condition $F_p(\xi) = G_p(-i\xi)$ at $\xi \to 0$. As $\xi \to \infty$, the bounce solution $\phi_B(\xi)$ undergoes damped oscillations around the true vacuum $\phi = \phi_T$. Therefore at $\xi \to \infty$, $F_p(\xi)$ will generally have the form,

$$F_p(\xi) = \frac{2}{\sqrt{\pi}}\overline{c_{1p}}\overline{f_p(\xi)} + \frac{2}{\sqrt{\pi}}\overline{c_{2p}}f_p(\xi)\,, \tag{2.26}$$

where

$$f_p(\xi) = \frac{\sqrt{\pi}}{2}e^{\pi p/2}H_{ip}^{(2)}(M\xi);\quad M^2 = V''(\phi_T). \tag{2.27}$$

Note that $f_p(\xi)$ is the positive frequency function for the Minkowski vacuum [13]. That is, one can expand the field operator as

$$\varphi(\xi,\chi,\Omega) = \int_0^\infty dp \sum_{\ell m}\left[\frac{f_p(\xi)}{\xi}Y_{p\ell m}(\chi,\Omega)a_{p\ell m} + \frac{\overline{f_p(\xi)}}{\xi}\overline{Y_{p\ell m}(\chi,\Omega)}a_{p\ell m}^\dagger\right]\,, \tag{2.28}$$

and the Minkowski vacuum is annihilated by $a_{p\ell m}$. On the other hand, as mentioned previously, the quantum state after tunneling is a "vacuum" state $|\tilde{0}\rangle$ specified by regarding $\overline{F_p(\xi)}$ as the (unnormalized) "positive frequency" functions. The orthonormalized positive frequency mode functions are then given by $u_{p\ell m}$ with $\overline{F_p}$ replaced by

$$U_p(\xi) = \sqrt{\frac{\pi}{4(|c_{1p}|^2 - |c_{2p}|^2)}}\overline{F_p(\xi)}\,. \tag{2.29}$$

Thus the spectrum of created particles $n_p$ observed by the Minkowski vacuum observer in the asymptotically future region is given by

$$\langle\tilde{0}|a_{p\ell m}^\dagger a_{p'\ell m}|\tilde{0}\rangle = n_p\delta(p-p');\quad n_p = \frac{1}{\left|c_{1p}/c_{2p}\right|^2 - 1}\,. \tag{2.30}$$

FIG. 2. The potential of the tunneling field.

Here we need to comment on the scaling property of this system. Under the rescaling given by

$$\phi \to g\phi, \quad \phi_i \to g\phi_i \quad (i = 1, 2)$$
$$x \to g'x,$$
$$m \to m/g', \quad \mu \to \mu/g', \quad M \to M/g', \tag{3.4}$$

the action scales as

$$S \to \left(\frac{g}{g'}\right)^2 S. \tag{3.5}$$

Thus the system is transformed to the same system but with a different Planck constant. Since the decay rate of the false vacuum is determined by the value of the action of the bounce solution, the tunneling rate changes under this rescaling. However as the field equation is invariant, the particle creation rate does not change. Thus every rescaled model is equivalent with each other as long as our interest is restricted to the particle creation. Therefore specifying the only three non-dimensional parameters, say $m/M$, $\mu/M$ and $\phi_2/\phi_1$ is sufficient for our purpose.

### A. bounce solution

First let us consider the bounce solution. In our potential model, we may regard the bubble wall to be the spacetime region surrounded by the spheres at $\xi_1$ and $\xi_2$ where $\phi_i = \phi_B(\xi_i)$ ($i = 1, 2$). So let us call $\xi_1$ and $\xi_2$ the inner and outer radii of the wall, respectively. With this definition, the wall is nothing but the negative mass-squared region. Here we tacitly assumed that the wall is entirely contained in the Euclidean region $\mathcal{E}$; $0 < \xi_1 < \xi_2 < \infty$. But there exists a case in which the wall extends to the Lorentzian region $\mathcal{M}$. Since the scalar field $\phi_B$ can get over the potential barrier only in $\mathcal{E}$, the outer edge of the wall $\xi_2$ must be in $\mathcal{E}$, but the inner edge of the wall $\xi_1$ can be either in $\mathcal{E}$ or in $\mathcal{M}$. We call the former the EE case, while the latter the EL case. We first consider the EE case. The EL case will be discussed at the end of this subsection.

For the potential (3.1), Eq.(2.19) reduces to Bessel's differential equation. Therefore the bounce solution $\phi_B(\xi_E)$ which satisfies the boundary condition (2.20) takes the form,

$$\frac{\phi_B(\xi_E)}{\phi_1} = \begin{cases} A(M\xi_E)^{-1} I_1(M\xi_E) \,; & 0 \leq \xi_E < \xi_1, \\ B_1(\mu\xi_E)^{-1} J_1(\mu\xi_E) + B_2(\mu\xi_E)^{-1} N_1(\mu\xi_E) + \dfrac{a}{\mu^2} \,; & \xi_1 \leq \xi_E < \xi_2, \\ C(m\xi_E)^{-1} K_1(m\xi_E) - \dfrac{c}{m^2} \,; & \xi_2 \leq \xi_E < \infty, \end{cases} \tag{3.6}$$

where $J_n$ and $N_n$ are the Bessel functions of the first and second kinds, respectively, and $I_n$ and $K_n$ are the modified Bessel functions of the first and second kinds, respectively. There are six unknown (non-dimensional) variables in the above; $A$, $B_1$, $B_2$, $C$, $M\xi_1$ and $M\xi_2$.

As the potential is constructed to be smooth to its first derivative, $\phi_B$, $d\phi_B/d\xi_E$ and $d^2\phi_B/d\xi_E^2$ must be continuous everywhere. Requiring this continuity at the junction points $\xi_E = \xi_i$, together with the equalities $\phi_i = \phi_B(\xi_i)$ ($i = 1, 2$), leads to the following algebraic equations:

$$\begin{aligned} 1 &= A(M\xi_1)^{-1} I_1(M\xi_1) = (\mu\xi_1)^{-1} \left[ B_1 J_1(\mu\xi_1) + B_2 N_1(\mu\xi_1) \right] + \frac{a}{\mu^2}, \\ A I_2(M\xi_1) &= -B_1 J_2(\mu\xi_1) - B_2 N_2(\mu\xi_1), \\ A I_3(M\xi_1) &= \frac{\mu}{M} \left[ B_1 J_3(\mu\xi_1) + B_2 N_3(\mu\xi_1) \right], \end{aligned} \tag{3.7}$$

and

$$\begin{aligned} \frac{\phi_2}{\phi_1} &= C(m\xi_2)^{-1} K_1(m\xi_2) - \frac{c}{m^2} = (\mu\xi_2)^{-1} \left[ B_1 J_1(\mu\xi_2) + B_2 N_1(\mu\xi_2) \right] + \frac{a}{\mu^2}, \\ C K_2(m\xi_2) &= B_1 J_2(\mu\xi_2) + B_2 N_2(\mu\xi_2), \\ C K_3(m\xi_2) &= \frac{\mu}{m} \left[ B_1 J_3(\mu\xi_2) + B_2 N_3(\mu\xi_2) \right], \end{aligned} \tag{3.8}$$

where the second equality in the first line of each set of the equations is the continuity of $\phi_B$, the second line is that of $d\phi_B/d\xi_E$, and the third line is that of $d/d\xi_E(\xi_E^{-1} d\phi_B/d\xi_E)$. Of course, these equations are not independent with each other; the third lines of each set can be derived from the rest of equations. Hence there are six independent equations which are necessary and sufficient to determine $A$, $B_1$, $B_2$, $C$, $M\xi_1$ and $M\xi_2$.

The analytic continuation of this bounce solution to $\mathcal{M}$ is given by the replacement of $\xi_E$ by $-i\xi$ as before:

$$\phi_B(-i\xi) = A(M\xi)^{-1} J_1(M\xi) \phi_1 \quad . \tag{3.9}$$

As the oscillation around the true vacuum attenuates, $\phi_B$ does not reach the junction point $\phi_1$ any more in $\mathcal{M}$. Hence $\phi_B$ is confined in the region $V''(\phi) = M^2$ and there is no additional particle creation after tunneling in this case.

Here it is to be noted that the equation for the mode function $G_p(\xi_E)$ (2.24) is expressed not in terms of the original potential parameters $m/M$, $\mu/M$ and $\phi_2/\phi_1$ but rather of parameters which specify the wall configuration. In the present case, the wall configuration parameters are $m/M$, $\mu/M$, $M\xi_1$ and $M\xi_2$. Thus instead of the original potential parameters we may regard these four parameters as those specify the model, three of which can be given freely. A convenient choice for the discussion given below is to give $m/M$, $M\xi_1$, $M\xi_2$. With this choice we find $\mu/M$ is determined from the four equations consisted of the second and third lines of Eqs. (3.7) and (3.8), despite the fact that there are five unknowns; $A$, $B_1$, $B_2$, $C$ and $\mu/M$. This is because these four equations are homogeneous with respect to $A$, $B_1$, $B_2$ and $C$. In fact, from the condition that there exists a non-trivial solution for $A$, $B_1$, $B_2$ and $C$, we can derive the equation to determine $\mu/M$,

$$\det \begin{pmatrix} I_3(M\xi_1)J_2(\mu\xi_1) + \dfrac{\mu}{M}I_2(M\xi_1)J_3(\mu\xi_1) & I_3(M\xi_1)N_2(\mu\xi_1) + \dfrac{\mu}{M}I_2(M\xi_1)N_3(\mu\xi_1) \\ K_3(m\xi_2)J_2(\mu\xi_2) - \dfrac{\mu}{m}K_2(m\xi_2)J_3(\mu\xi_2) & K_3(m\xi_2)N_2(\mu\xi_2) - \dfrac{\mu}{m}K_2(m\xi_2)N_3(\mu\xi_2) \end{pmatrix} = 0. \tag{3.10}$$

The fact that $\mu/M$ is determined in this manner can be explained by taking the derivative of the equation for the bounce (2.19),

$$\left[ -\frac{d^2}{d\xi_E^2} + \frac{15}{4\xi_E^2} + V''(\xi_E) \right] \psi = 0, \tag{3.11}$$

where $\psi := \xi_E^{3/2} d\phi_B/d\xi_E$. The function $\psi$ satisfies the boundary condition,

$$\psi \to 0 \quad \text{for} \quad \xi_E \to 0 \text{ and } \infty, \tag{3.12}$$

and it has no node. This is just the condition that the ground state wave function for a one-dimensional quantum mechanics with the potential $15/(4\xi_E^2) + V''(\xi_E)$ should have zero energy eigenvalue. Given $m/M$, $M\xi_1$ and $M\xi_2$, this condition is exactly what is expressed by Eq.(3.10). Moreover, from the above analogy with the quantum mechanics it is clear that $\mu$ cannot exceed the order of $1/(\xi_2 - \xi_1)$.

Now let us discuss the EL case in which the inner edge of the wall is in the Lorentzian region $\mathcal{M}$. In this case the bounce solution $\phi_B$ is given by

$$\frac{\phi_B(\xi_E)}{\phi_1} = \begin{cases} B(\mu\xi_E)^{-1} J_1(\mu\xi_E) + \dfrac{a}{\mu^2}; & 0 \le \xi_E < \xi_2, \\ C(m\xi_E)^{-1} K_1(m\xi_E) - \dfrac{c}{m^2}; & \xi_2 \le \xi_E < \infty, \end{cases} \tag{3.13}$$

with the junction conditions,

$$\frac{\phi_2}{\phi_1} = C(m\xi_2)^{-1} K_1(m\xi_2) - \frac{c}{m^2} = (\mu\xi_2)^{-1} B J_1(\mu\xi_2) + \frac{a}{\mu^2},$$
$$CK_2(m\xi_2) = BJ_2(\mu\xi_2),$$
$$CK_3(m\xi_2) = \frac{\mu}{m} B J_3(\mu\xi_2). \tag{3.14}$$

By the same reason as in the EE case, we have three independent equations and there are the same number of unknowns; $B, C$ and $M\xi_2$.

The analytic continuation of this bounce solution to $\mathcal{M}$ is

$$\frac{\phi_B(-i\xi)}{\phi_1} = \begin{cases} B(\mu\xi)^{-1} I_1(\mu\xi) + \dfrac{a}{\mu^2}; & 0 \le \xi < \xi_1, \\ A_1(M\xi)^{-1} J_1(M\xi) + A_2(M\xi)^{-1} N_1(M\xi); & \xi_1 \le \xi < \infty, \end{cases} \tag{3.15}$$

with the junction conditions,

$$1 = B(\mu\xi_1)^{-1} I_1(\mu\xi_1) + \frac{a}{\mu^2} = A_1(M\xi_1)^{-1} J_1(M\xi_1) + A_2(M\xi_1)^{-1} N_1(M\xi_1),$$
$$BI_2(\mu\xi_1) = A_1 J_2(M\xi_1) + A_2 N_2(M\xi_1),$$
$$BI_3(\mu\xi_1) = \frac{M}{\mu} \left( A_1 J_3(M\xi_1) + A_2 N_3(M\xi_1) \right). \tag{3.16}$$

Here again only three of the above equations are independent and the unknown parameters are $A_1$, $A_2$ and $M\xi_1$, assuming $B$ is known by solving Eqs.(3.14).

As in the EE case, we may choose the wall configuration parameters, $m/M$, $M\xi_1$ and $M\xi_2$, as independent model parameters. As before, $\mu/M$ is determined from the last two equations of (3.14). Specifically the equation to be solved is

$$K_3(m\xi_2)J_2(\mu\xi_2) = \frac{\mu}{m}K_2(m\xi_2)J_3(\mu\xi_2). \tag{3.17}$$

Thus in the EL case $\mu/m$ is independent of $M\xi_1$ and is given as a function of $m\xi_2$. More precisely, a close analysis of Eq. (3.17) reveals that $\mu\xi_2$ varies from $j_{1,1} \sim 3.8132$ to $j_{2,1} \sim 5.1356$ as $m\xi_2$ varies from zero to infinity, where $j_{m,n}$ is the $n$-th zero point of the Bessel function $J_m(z)$.

### B. mode functions and particle spectrum

As we have found the background solution $\phi_B$, we now turn to the equation for the mode function (2.24). As in the case of the bounce solution, Eq. (2.24) reduces again to Bessel's differential equation. First consider the EE case. In this case, $G_p(\xi_E)$ in $\mathcal{E}$ is solved to give

$$G_p(\xi_E) = \begin{cases} a_{1p}K_{ip}(M\xi_E) + a_{2p}I_{ip}(M\xi_E); & 0 \leq \xi_E < \xi_1, \\ b_{1p}J_{ip}(\mu\xi_E) + b_{2p}N_{ip}(\mu\xi_E); & \xi_1 \leq \xi_E < \xi_2, \\ K_{ip}(m\xi_E); & \xi_2 \leq \xi_E < \infty, \end{cases} \tag{3.18}$$

where the function $K_{ip}$ is chosen in the interval $\xi_2 \leq \xi_E < \infty$ in accordance with the boundary condition (2.25). This time, $G_p(\xi_E)$ and its first derivative are necessarily continuous at the junction points. These conditions determine the coefficients $a_{1p}$, $a_{2p}$, $b_{1p}$ and $b_{2p}$. Using the matrix notation, $a_{1p}$ and $a_{2p}$ are expressed as

$$\begin{pmatrix} a_{1p} \\ a_{2p} \end{pmatrix} = \mathbf{X}_p \mathbf{Y}_p \begin{pmatrix} K_{ip}(m\xi_2) \\ mK'_{ip}(m\xi_2) \end{pmatrix}, \tag{3.19}$$

where

$$\mathbf{X}_p = \xi_1 \begin{pmatrix} MI'_{ip}(M\xi_1) & -I_{ip}(M\xi_1) \\ -MK'_{ip}(M\xi_1) & K_{ip}(M\xi_1) \end{pmatrix} \begin{pmatrix} J_{ip}(\mu\xi_1) & N_{ip}(\mu\xi_1) \\ \mu J'_{ip}(\mu\xi_1) & \mu N'_{ip}(\mu\xi_1) \end{pmatrix}, \tag{3.20}$$

$$\mathbf{Y}_p = \frac{\pi\xi_2}{2} \begin{pmatrix} \mu N'_{ip}(\mu\xi_2) & -N_{ip}(\mu\xi_2) \\ -\mu J'_{ip}(\mu\xi_2) & J_{ip}(\mu\xi_2) \end{pmatrix}. \tag{3.21}$$

The mode function $F_p(\xi)$ in the Lorentzian region $\mathcal{M}$ is given by Eq. (2.26), where the coefficients $\overline{c_{1p}}$ and $\overline{c_{2p}}$ are related to $a_{1p}$ and $a_{2p}$ as follows. Following the prescription $\xi_E \to -i\xi$, analytic continuation of the Bessel functions gives

$$K_{ip}(-iM\xi) = \frac{\pi i}{2}e^{-\pi p/2}H^{(1)}_{ip}(M\xi),$$

$$I_{ip}(-iM\xi) = \frac{1}{2}e^{\pi p/2}\left(H^{(1)}_{ip}(M\xi) + H^{(2)}_{ip}(M\xi)\right). \tag{3.22}$$

This implies the following relation,

$$\begin{pmatrix} \overline{c_{1p}} \\ \overline{c_{2p}} \end{pmatrix} = \frac{1}{2}\begin{pmatrix} \pi i & e^{\pi p} \\ 0 & 1 \end{pmatrix}\begin{pmatrix} a_{1p} \\ a_{2p} \end{pmatrix}. \tag{3.23}$$

Since

$$G_p(\xi_E) = \left(a_{1p} - \frac{i}{\pi}\sinh(\pi p)a_{2p}\right)K_{ip}(M\xi_E) + \frac{a_{2p}}{2}\left[I_{ip}(M\xi_E) + I_{-ip}(M\xi_E)\right], \tag{3.24}$$

for $0 \leq \xi_E < \xi_1$ where both $K_{ip}$ and $I_{ip}+I_{-ip}$ are real, the fact that $G_p(\xi_E)$ is real in $\mathcal{E}$ implies both $a_{1p} - \frac{i}{\pi}\sinh(\pi p)a_{2p}$ and $a_{2p}$ must be real. Using this property, the expression (2.30) for the particle spectrum $n_p$ is rewritten as

$$n_p = \frac{a_{2p}^2}{\pi^2 |a_{1p}|^2} \quad , \tag{3.25}$$

where $a_{1p}$ and $a_{2p}$ are given by Eq. (3.19). This expression, although exact, is too complicated that it is almost impossible to gain any physical insight from it. In the next section, we will examine several extreme situations in which we can obtain approximate analytical formulas which are more comprehensible. We will also evaluate Eq.(3.25) numerically to fill the parameter regions not covered by the approximate formulas.

Now we consider the EL case. In this case, the mode function $G_p(\xi_E)$ and its analytic continuation $F_p(\xi) = G_p(-i\xi)$ are expressed as

$$G_p(\xi_E) = \begin{cases} K_{ip}(m\xi_E); & \xi_2 \leq \xi_E < \infty, \\ b_{1p} J_{ip}(\mu\xi_E) + b_{2p} N_{ip}(\mu\xi_E); & 0 \leq \xi_E < \xi_2, \end{cases} \tag{3.26}$$

$$F_p(\xi) = \begin{cases} \tilde{b}_{1p} I_{ip}(\mu\xi) + \tilde{b}_{2p} K_{ip}(\mu\xi); & 0 \leq \xi < \xi_1, \\ \overline{c_{1p}} e^{-\pi p/2} H_{ip}^{(1)}(M\xi) + \overline{c_{2p}} e^{\pi p/2} H_{ip}^{(2)}(M\xi); & \xi_1 \leq \xi < \infty. \end{cases} \tag{3.27}$$

Through the origin $\xi_E = \xi = 0$, the Bessel functions are analytically continued as

$$J_{ip}(-i\mu\xi) = e^{\pi p/2} I_{ip}(\mu\xi) \quad ,$$
$$N_{ip}(-i\mu\xi) = -i e^{\pi p/2} I_{ip}(\mu\xi) - \frac{2}{\pi} e^{-\pi p/2} K_{ip}(\mu\xi) \quad . \tag{3.28}$$

With the aid of these relations, we get

$$\begin{pmatrix} \overline{c_{1p}} \\ \overline{c_{2p}} \end{pmatrix} = \mathbf{Z}_p \mathbf{W}_p \mathbf{Y}_p \begin{pmatrix} K_{ip}(m\xi_2) \\ m K'_{ip}(m\xi_2) \end{pmatrix} , \tag{3.29}$$

where

$$\mathbf{Z}_p = \frac{\pi\xi_1}{-4i} \begin{pmatrix} M e^{\pi p/2} H^{(2)\prime}_{ip}(M\xi_1) & -e^{\pi p/2} H^{(2)}_{ip}(M\xi_1) \\ -M e^{-\pi p/2} H^{(1)\prime}_{ip}(M\xi_1) & e^{-\pi p/2} H^{(1)}_{ip}(M\xi_1) \end{pmatrix} \begin{pmatrix} I_{ip}(\mu\xi_1) & K_{ip}(\mu\xi_1) \\ \mu I'_{ip}(\mu\xi_1) & \mu K'_{ip}(\mu\xi_1) \end{pmatrix} , \tag{3.30}$$

$$\mathbf{W}_p = \begin{pmatrix} e^{\pi p/2} & -i e^{\pi p/2} \\ 0 & -\frac{2}{\pi} e^{-\pi p/2} \end{pmatrix} , \tag{3.31}$$

and $\mathbf{Y}_p$ is given in Eq. (3.21). Substituting $\overline{c_{1p}}$ and $\overline{c_{2p}}$ given by Eq. (3.29) into Eq. (2.30), we obtain the expression for the spectrum of the created particles in the EL case.

## IV. DETAILED ANALYSIS OF PARTICLE SPECTRUM

As shown in [13], the spectrum of created particles has the general feature that $n_p$ is nearly constant for $0 \leq p \lesssim 1$ and decreases exponentially as $e^{-2\pi p}$ for $p \gg 1$. Hence the particle spectrum is basically obtained if $n_0 := n_{p=0}$ is known. Note that $p = 1$ corresponds to the curvature scale of the $\xi = const.$ hypersurface in the Milne universe. It should be also noted $p = 0$ does not corresponds to the zero mode of the $\xi = const.$ hypersurface but to the mode with characteristic scale of the curvature as well. Furthermore, if one considers implications of the present analysis to the one-bubble inflation scenario, what one wants to know most is the curvature perturbation spectrum on large scales comparable to the spatial curvature scale, which is described by $n_p$ at $p \lesssim 1$. Thus we may focus on the plateau of $n_p$ at $p \lesssim 1$. For definiteness, we take $n_0$ as the representative value.

For limiting cases in which the argument of the (modified) Bessel functions is very small or large compared to unity, we can derive a rather simple analytical expression for $n_0$ by using the asymptotic behaviour of the (modified) Bessel functions which may be found in Ref. [16]. As discussed in the previous section, there are two ways of specifying parameters of the model. The potential parameters are less directly related to $n_p$ than the wall configuration parameters. Therefore, in what follows, we first consider cases with extreme values of the wall configuration parameters. After disclosing the relationship between the wall configuration parameters and the particle spectrum, we then interpret the results in terms of the original potential parameters by analyzing the relations between these two sets of parameters.

FIG. 3. The typical shape of the particle spectrum $n_p$.

### A. $n_0$ as a function of wall configuration parameters

(i) Thin wall EE case:

We first consider the case when $\tilde{m}\xi_1$ and $\tilde{m}\xi_2 \to \infty$, where $\tilde{m}$ represents either $m$, $M$ or $\mu$, and $(\xi_2 - \xi_1)/\xi_1 \ll 1$, which corresponds to the thin wall limit. We regard $1/\tilde{m}\xi_1$ and $1/\tilde{m}\xi_2$ as small parameters, which we denote by $\epsilon$, and assume $\mu(\xi_2 - \xi_1) \lesssim 1$.

After tedious manipulations, the expansion of Eq. (3.10) with respect to $\epsilon$ to $O(\epsilon)$ gives

$$(Mm - \mu^2)\mathcal{T} + \mu(M + m) - \frac{15}{8\xi_1}\left(\frac{M}{\mu} + \frac{\mu}{M}\right)(m - \mu\mathcal{T}) + \frac{15}{8\xi_2}\left(\frac{m}{\mu} + \frac{\mu}{m}\right)(M - \mu\mathcal{T}) = O\left(\epsilon^2\right), \tag{4.1}$$

where

$$\mathcal{T} := \tan\mu(\xi_2 - \xi_1). \tag{4.2}$$

With the aid of this relation, the particle spectrum at $p = 0$ given by (3.25) is evaluated as

$$n_0 \sim \left[\left(\frac{M}{\mu} + \frac{\mu}{M}\right)\left(\frac{1}{\mu\xi_1} - \frac{1}{\mu\xi_2}\right)\right]^{-2} \exp(-4M\xi_1). \tag{4.3}$$

Equivalently, again using Eq.(4.1), $n_0$ may be expressed as

$$n_0 \sim \left[\left(\frac{M}{\mu} + \frac{\mu}{M}\right)\arctan\left(\frac{\mu(M + m)}{\mu^2 - Mm}\right)\right]^{-2} (\mu\xi_1)^4 \exp(-4M\xi_1), \tag{4.4}$$

where arctan takes the value between 0 and $\pi$.

One sees that $n_0$ is exponentially suppressed as $e^{-4M\xi_1}$, the feature that has been found in the previous analysis of the thin wall limit [13]. However, one also finds a large factor $(\mu\xi_1)^4$ in front, which were absent in the simple model discussed in [13].

(ii) Boundary between the EE and EL cases:

The boundary between the EE and EL cases is given by the limit $M\xi_1 \to 0$ and $\mu\xi_1 \to 0$. In this case the limit of either $m\xi_2 \to 0$ or $m\xi_2 \to \infty$ can be treated analytically.

As mentioned below Eq. (3.17), for $m\xi_2 \to 0$, $\mu$ is fixed as

$$\mu\xi_2 = j_{1,1} \sim 3.8132. \tag{4.5}$$

Then we obtain

$$n_0 \sim \frac{1}{4}\left(k_1 + \frac{2}{\pi}\ln\left(\frac{M}{\mu}\right)\right)^2, \tag{4.6}$$

where $k_1$ is a numerical factor given by

$$k_1 = \left\{ j_{1,1} N_1(j_{1,1}) \left( \ln\left(\frac{m\xi_2}{2}\right) + \gamma \right) + N_0(j_{1,1}) \right\} \Big/ J_0(j_{1,1})$$
$$\sim -3.9245 \ln\left(\frac{m\xi_2}{2}\right) - 2.3929, \tag{4.7}$$

and $\gamma \sim 0.5772$ is the Euler constant.

For $m\xi_2 \to \infty$, $\mu$ is fixed as

$$\mu\xi_2 = j_{2,1} \sim 5.1356, \tag{4.8}$$

Then we obtain

$$n_0 \sim \frac{1}{4}\left(k_2 + \frac{2}{\pi}\ln\left(\frac{M}{\mu}\right)\right)^2, \tag{4.9}$$

where $k_2$ is given by

$$k_2 = \frac{N_0(j_{2,1})}{J_0(j_{2,1})} \sim 2.4602. \tag{4.10}$$

Thus, in both of the limiting cases, $n_0$ is of order unity unless the ratio $M/\mu$ or $m/\mu$ becomes too large or too small.

(iii) EL case with $\mu\xi_1 \to \infty$:

In the EL case with $\mu\xi_1 \to \infty$ and $M\xi_1 \to \infty$, a simple expression for $n_0$ can be obtained for $m\xi_2 \to \infty$ or $m\xi_2 \to 0$.

First consider the limit $m\xi_2 \to 0$. In this case $\mu$ is given also by Eq. (4.5). After a straightforward calculation, we then find

$$n_0 \sim \frac{k_1^2 + 1}{16} \frac{\mu^2 + M^2}{\mu M} \exp(2\mu\xi_1). \tag{4.11}$$

In the case of the limit $m\xi_2 \to \infty$, $\mu$ is given by Eq. (4.8). Then we obtain

$$n_0 \sim \frac{k_2^2 + 1}{16} \frac{\mu^2 + M^2}{\mu M} \exp(2\mu\xi_1). \tag{4.12}$$

Thus in both limits of $m\xi_2$, the particle spectrum has the exponential factor $e^{2\mu\xi_1}$. This suggests that it is a common factor for the EL case irrespective of the parameters.

Summarizing the above results of analytical tractable cases, we expect that the gross dependence of $n_0$ on the wall configuration parameters is

$$n_0 \sim \exp(-4M\xi_1) \quad , \tag{4.13}$$

for the EE case, and

$$n_0 \sim \exp(2\mu\xi_1) \quad , \tag{4.14}$$

for the EL case.

In order to test our expectation mentioned above, we have numerically evaluated $n_0$ for various values of the wall configuration parameters. Figures 4 and 5 show $n_0/\exp(-4M\xi_1)$ for the EE case and $n_0/\exp(2\mu\xi_1)$ for the EL case, respectively, as functions of the parameters $M\xi_1$ and $M\xi_2$, for typical ratios of $m/M$; $m/M = 0.1$, 1 and 10. We see that the approximations as Eqs. (4.13) and (4.14) are better than one might have anticipated except for some special cases.

(c)

FIG. 4. The contour plots of the particle creation rate $n_{p=0}$ versus $M\xi_1$ and $M\xi_2 - M\xi_1$ in the EE case. They are plotted for a few typical values of $m/M$, i.e., (a) $m/M = 0.1$, (b) $m/M = 1$ and (c) $m/M = 10$.

(c)

FIG. 5. The contour plots of $n_{p=0}$ versus $M\xi_1$ and $M\xi_2$ in the EL case. As before they are plotted for the same typical values of $m/M$.

As for the EE case, we see a ditch in the figure, but it is in some sense superficial. As is also observed in Fig. 3, there

FIG. 6.

FIG. 7. The plot of the function $f(x)$.

FIG. 8.

If we release a particle from the same point $\phi_I$ as in the original potential for the case of the potential $\widetilde{V}$, it cannot reach the false vacuum $\widetilde{\phi}_F$. In order to make it reach $\widetilde{\phi}_F$, the particle must be released from a point $\phi = \widetilde{\phi}_I$ closer to the true vacuum, i.e., $\widetilde{\phi}_I > \phi_I$. Then $\xi_1$, which is the value of $x_E$ at $\phi_1$, must become larger. Thus we conclude $M\widetilde{\xi}_1 > M\xi_1$ for $\widetilde{\phi_2/\phi_1} > \phi_2/\phi_1$ with fixed $m/M$ and $\mu/M$.

In contrast to the EL case, we were unable to find a simple function of the potential parameters which is directly related to the amplitude of particle creation. This is because the bounce solution in the EE case depends on the mass at the true vacuum $M$ as well as $m$ and $\mu$. Therefore $M\xi_1$ is generally fully dependent on all the three potential parameters. However, it is still possible to relate the shape of the potential to the particle creation in a couple of limiting cases, which we will discuss below.

As noted in the beginning of section 3, $\phi_2/\phi_1$ is bounded from above by the condition that $V(\phi_F) > V(\phi_T)$. For $\phi_2/\phi_1$ close to the maximum value, we have $M\xi_1 \gg 1$, which is just the thin-wall limit. In this limit, the particle creation is exponentially suppressed as $e^{-4M\xi_1}$. In other word, as the wall radius becomes larger, the quantum state inside a nucleated bubble becomes closer to the Minkowski vacuum state. In this case, the wall radius is known to be given by [7]

$$\xi_1 \sim \xi_2 \sim 3S_1/\Delta V, \qquad (4.23)$$

where

$$S_1 := \int_{\phi_F}^{\phi_T} d\phi \sqrt{2V(\phi)}, \qquad (4.24)$$

and

$$\Delta V := V(\phi_F) - V(\phi_T). \qquad (4.25)$$

Thus the amplitude of particle creation $n_0$ is approximately given by

$$n_0 \sim \exp\left(-\frac{12MS_1}{\Delta V}\right). \qquad (4.26)$$

On the other hand, $\phi_2/\phi_1$ is bounded also from below in order for the potential barrier to exist. Now as $\phi_2/\phi_1$ decreases, $\xi_1$ decreases and before the potential barrier vanishes $\xi_1$ becomes zero, which is just the boundary of the EE case and the EL case. Thus in this limit, the analysis of the $\xi_1 \to 0$ limit of the EL case is also appropriate.

Summarizing the above analyses for the EL and EE cases, we conclude that $n_0$ is approximately determined in terms of the potential parameters as

$$n_0 \sim \begin{cases} \mathcal{F}_\phi; & \mathcal{F}_\phi \gtrsim 1, \\ \exp(-12MS_1/\Delta V); & \mathcal{F}_\phi \ll 1. \end{cases} \qquad (4.27)$$

Before closing this section, one comment is in order. In this paper, we have considered a potential model with a constant mass around the true vacuum. However in order for our results to be valid, the mass $M(\xi)$ on the true vacuum side is not necessarily strictly constant. The only restriction is that $M(\xi)$ should vary sufficiently slowly. Namely if

$$\frac{dM(\xi)}{d\xi} \Big/ M(\xi) \ll M(\xi) \quad , \qquad (4.28)$$

the mode function evolves adiabatically and there will be no additional particle creation on the true vacuum side. Now in a simple version of the one-bubble inflation scenario, the tunneling field also plays the role of the inflaton field inside the bubble, and the mass of the inflaton field changes very slowly in the slow rolling phase. Thus our results are expected to give non-trivial implications to the one-bubble inflation scenario, at least the effect of tunneling to the spectrum of the inflaton field fluctuations is concerned. We plan to make a detailed investigation of this issue in a future publication.

## V. DISCRETE MODES

So far, we have not carefully investigated the completeness and normalization of the mode functions for the description of a quantum state of the field $\varphi$. In order to specify a quantum state completely, we need a set of all possible mode functions which have properly normalized Klein-Gordon inner products on a Cauchy surface. However, the spacetime inside the forward light cone of the center of a bubble, which is described by the Milne universe, does not contain any Cauchy surface of the whole Minkowski space.

Thus we first need to introduce new coordinates which cover the spacelike region outside the forward light cone and which respect the symmetry of the bounce solution. Such coordinates are known as the (spherical) Rindler coordinates and the spacetime covered by them is called the Rindler space. The metric of the Rindler space is given by

$$ds^2 = d\xi_R^2 + \xi_R^2 \left(-d\chi_R^2 + \cosh^2 \chi_R d\Omega_{(2)}^2\right), \qquad (5.1)$$

where the $\chi_R = const.$ hypersurface of the Rindler space is a Cauchy surface of the whole Minkowski space. These coordinates are related to the Milne coordinates as

$$\xi_R = e^{-\pi i/2}\xi = \xi_E, \quad \chi_R = \chi + \frac{\pi i}{2}. \qquad (5.2)$$

Now we have to evaluate the Klein-Gordon norms of the mode functions $u_{p\ell m}(\xi,\chi,\Omega)$ in the Rindler space. In order to do so, we need the analytic continuation of $\overline{u_{p\ell m}}$ to the Rindler space:

$$\overline{u_{p\ell m}} = \frac{G_p(\xi_R)}{\xi_R}\overline{v_{p\ell}(\chi_R)Y_{\ell m}(\Omega)}, \tag{5.3}$$

where

$$\overline{v_{p\ell}(\chi_R)} = f_{p\ell}(\chi_R - i\pi/2) = \left|\frac{\Gamma(ip+\ell+1)\Gamma(-ip+\ell+1)}{\Gamma(ip)\Gamma(-ip)}\right|^{1/2}\frac{1}{\sqrt{-i\cosh\chi_R}}P_{ip-1/2}^{-\ell-1/2}(-i\sinh\chi_R). \tag{5.4}$$

We note that $v_{p\ell}(\chi_R)$ now plays the role of a positive frequency function for the state $|\tilde{0}\rangle$. In the same way as presented in Appendix A of Ref. [15], the Klein-Gordon inner product of $u_{p\ell m}$ on the $\chi_R = const.$ hypersurface can be evaluated as

$$\langle u_{p\ell m}, u_{p'\ell'm'}\rangle := i\cosh^2\chi_R \int_0^\infty d\xi_R \xi_R \left(\frac{\partial u_{p\ell m}}{\partial \chi_R}\overline{u_{p'\ell'm'}} - u_{p\ell m}\frac{\partial \overline{u_{p'\ell'm'}}}{\partial \chi_R}\right)$$
$$= N_{p\ell}^{(1)} N_{pp'}^{(2)} \delta_{\ell\ell'}\delta_{mm'}, \tag{5.5}$$

where

$$N_{p\ell}^{(1)} = i\cosh^2\chi_R\left(\frac{\partial v_{p\ell}}{\partial\chi_R}\overline{v_{p\ell}} - v_{p\ell}\frac{\partial \overline{v_{p\ell}}}{\partial\chi_R}\right), \quad N_{pp'}^{(2)} = \int_0^\infty \frac{d\xi_R}{\xi_R}\overline{G_p(\xi_R)}G_{p'}(\xi_R). \tag{5.6}$$

In the above, we have put $p' = p$ when evaluating $N_{p\ell}^{(1)}$, anticipating that the factor $N_{pp'}^{(2)}$ should be zero if $p \neq p'$, which we will show below. The evaluation of the factor $N_{p\ell}^{(1)}$ is straightforward and we obtain

$$N_{p\ell}^{(1)} = \frac{2p}{\pi}\sinh \pi p. \tag{5.7}$$

For $p^2$ and $p'^2 > 0$ (in fact we have $p, p' > 0$), the factor $N_{pp'}^{(2)}$ can be evaluated from the behaviour of $G_p(\xi_R)$ near the origin alone. In the EE case, noting that Eq. (3.24) implies

$$G_p(\xi_R) = \frac{\overline{c_{1p}} - e^{-\pi p}\,\overline{c_{2p}}}{\sinh \pi p}I_{ip}(M\xi_R) - \frac{\overline{c_{1p}} - e^{\pi p}\,\overline{c_{2p}}}{\sinh \pi p}I_{-ip}(M\xi_R), \tag{5.8}$$

and using the fact that $i(c_{1p} - \cosh\pi p c_{2p})$ and $c_{2p}$ are real, we obtain

$$N_{pp'}^{(2)} = \frac{2\delta(p-p')}{p\sinh \pi p}\left(|c_{1p}|^2 - |c_{2p}|^2\right). \tag{5.9}$$

Thus we have

$$\langle u_{p\ell m}, u_{p'\ell'm'}\rangle = \frac{4\left(|c_{1p}|^2 - |c_{2p}|^2\right)}{\pi}\delta(p-p')\delta_{\ell\ell'}\delta_{mm'}. \tag{5.10}$$

This agrees with the norm calculated on the $\xi = const.$ hypersurface in the Milne universe, Eq. (2.29). We can show the same result holds also in the EL case. The reason why the Klein-Gordon inner product can be evaluated in the Milne universe, despite the fact that it does not contain any Cauchy surface, is due to the fast fall-off of the function $v_{p\ell}(\chi_R)$ with positive $p^2$ at $\chi_R \to \infty$. That is, the mode function $u_{p\ell m}$ vanishes fast enough at future null infinity of the Minkowski space.

In contrast, any mode function with negative $p^2$ has a divergent Klein-Gordon norm in the Milne universe. When the Klein-Gordon norm of a mode diverges, the normalized mode function vanishes. Hence it would not contribute to the quantum fluctuations of the field if it also vanishes fast enough at future null infinity. However, for $p^2 < 0$, the fall-off of a mode function at future null infinity is not fast enough. Thus, we cannot claim that the Klein-Gordon norm evaluated in the Milne universe is equivalent to that evaluated on a Cauchy surface. Therefore it sometimes occurs that some of the modes with negative $p^2$ do contribute to the quantum fluctuations of the field.

Specifically, as seen from Eq. (5.8), $G_p(\xi_R)$ with $p^2 < 0$ has the behavior $\sim \alpha_p \xi_R^{|p|} + \beta_p \xi_R^{-|p|}$ near the origin $\xi_R = 0$, where $\alpha_p$ and $\beta_p$ are determined by solving Eq. (2.24) with the boundary condition (2.25). This boundary condition

is a necessary condition for the normalizability also for the $p^2 < 0$ modes but not a sufficient one. For a mode to be normalizable, there is an additional condition that $\beta_p$ should vanish. For almost all the modes with negative $p^2$, $\beta_p$ do not vanish. Hence they do not contribute to the quantum fluctuations. However, for certain discrete sets of $p$, $\beta_p$ may happen to become zero. Then those modes become normalizable. Apparently, they are bound state modes and cannot be described in the particle picture.

Nevertheless, we can evaluate the contribution of these modes to the two-point function (Wightman function) which characterizes the quantum fluctuations of the field. Let us denote these discrete values of imaginary $p$ by $p_n$. Since the Wightman function is given by the summation of products of the mode functions, it can be divided into two parts as

$$G^+(x, x') = G_D^+(x, x') + G_C^+(x, x'). \tag{5.11}$$

The two parts $G_D^+$ and $G_C^+$ are the contributions from the discrete modes and the continuous modes, respectively. They are given by

$$\begin{aligned} G_D^+(x, x') &= \sum_{n\ell m} \frac{\pi}{2 p_n \sinh \pi p_n N_n^{(2)}} u_{p_n \ell m}(x) \overline{u_{p_n \ell m}(x')} \\ &= \frac{1}{2\pi^2} \sum_n \frac{\sinh |p_n| \zeta}{4\pi \sin \pi |p_n|} \frac{1}{N_n^{(2)}} \frac{G_{p_n}(\xi_R) G_{p_n}(\xi_R')}{\xi_R \xi_R'}, \end{aligned} \tag{5.12}$$

where

$$N_n^{(2)} := \int_0^\infty \frac{d\xi_R}{\xi_R} \left( G_{p_n}(\xi_R) \right)^2, \tag{5.13}$$

and

$$\begin{aligned} G_C^+(x, x') &= \int_0^\infty dp \sum_{\ell m} \frac{\pi}{4(|c_{1p}|^2 - |c_{2p}|^2)} u_{p\ell m}(x) \overline{u_{p\ell m}(x')} \\ &= \int_0^\infty dp \frac{1}{8\pi(|c_{1p}|^2 - |c_{2p}|^2)} \frac{p \sin p\zeta}{\sinh \zeta} \frac{\overline{G_p(\xi_R)} G_p(\xi_R')}{\xi_R \xi_R'}, \end{aligned} \tag{5.14}$$

where $\zeta$ is the Lorentz invariant distance between $x$ and $x'$,

$$\begin{aligned} \cosh \zeta &= \begin{cases} \cosh \chi \cosh \chi' - \sinh \chi \sinh \chi' \cos \Theta & \text{(in Milne universe)}, \\ -\sinh \chi_R \sinh \chi_R' + \cosh \chi_R \cosh \chi_R' \cos \Theta & \text{(in Rindler space)}, \end{cases} \\ \cos \Theta &= \cos \theta \cos \theta' + \sin \theta \sin \theta' \cos(\phi - \phi'). \end{aligned} \tag{5.15}$$

The second equalities of both Eqs. (5.12) and (5.14) are obtained after summation over $\ell$ and $m$, and they manifestly show the $O(3, 1)$ invariance.

In the present model, there exists at lease one series of such discrete modes. They are related to the perturbations of the wall location. The radial part of these mode functions is given by the derivative of the bounce solution as

$$G_{wall}(\xi_R) = \xi_R \frac{d\phi_B(\xi_R)}{d\xi_R}. \tag{5.16}$$

It is easy to show that this mode function satisfies Eq. (2.24) with $p^2 = -4$. If we put $p_n = 2i$ in Eq. (5.12) it diverges. However the divergence comes from the monopole ($\ell = 0$) and dipole ($\ell = 1$) parts of these modes, which just represent translations of the origin of the coordinates (so-called zero modes). In fact it is easy to check that these modes are represented by linear combinations of $d\phi_B/dx^\mu$ where $x^\mu$ are the usual Minkowski coordinates. Thus they should be removed. Then the remaining part of the Wightman function becomes finite but at the expense of losing the Lorentz invariance. Using the explicit form of the associated Legendre functions with special values of the indices,

$$P_{\nu-1/2}^{-1/2}(\cosh z) = \sqrt{\frac{2}{\pi \sinh z}} \frac{\sinh \nu z}{\nu}, \tag{5.17}$$

$$P_{\nu-1/2}^{-3/2}(\cosh z) = \sqrt{\frac{2}{\pi \sinh z}} \frac{\nu \cosh \nu z - \sinh \nu z \coth z}{\nu(\nu^2 - 1)},$$

the contribution to the Wightman function from these discrete modes is expressed as

$$G^+_{wall}(x,x') = \frac{\dot\phi_B(\xi_R)\dot\phi_B(\xi'_R)}{4\pi^2 N^{(2)}_{wall}}\left[\zeta\frac{\cosh 2\zeta}{\sinh\zeta} - \frac{11}{3}\cosh\zeta + \frac{8}{3}\sinh\chi_R\sinh\chi'_R\right], \quad (5.18)$$

where

$$N^{(2)}_{wall} = \int_0^\infty d\xi_R \xi_R \dot\phi_B^2(\xi_R). \quad (5.19)$$

The last term in the square brackets is not Lorentz invariant. However, since the removed monopole and dipole modes simply represent the global translation of the bubble, this apparent violation of the Lorentz invariance should not affect observable quantities. Focusing on the wall fluctuations in the Rindler space, this point has been discussed in detail by Garriga and Vilenkin [17].

In the Milne universe, one can interpret the fluctuation of the scalar field $\varphi$ as the perturbation of the intrinsic curvature of the $\phi = const.$ hypersurface. The scalar-type curvature perturbation is described by a single potential function $\mathcal{R}$ as [18]

$$\delta^{(3)}R^i_j = \frac{1}{\xi^2}\left[-\left(\nabla^i\nabla_j - \frac{1}{3}\delta^i_j\nabla^k\nabla_k\right) + \frac{4}{3}\delta^i_j(3 - \nabla^k\nabla_k)\right]\mathcal{R}(x)$$
$$= \frac{1}{\xi^2}\left[-(1+p^2)\mathcal{R}Y^i_j + \frac{4}{3}(4+p^2)\delta^i_j\mathcal{R}Y\right], \quad (5.20)$$

where $\nabla_i$ denotes the covariant derivative with respect to the metric on the unit 3-hyperboloid, and the second line is the harmonic expansion of the first line with $Y$ being the abbreviation for $Y_{p\ell m}(\chi,\Omega)$ and $Y^i_j$ being the traceless tensor harmonics defined by

$$Y^i_j = \frac{1}{1+p^2}\left[\nabla^i\nabla_j - \frac{1}{3}\delta^i_j\nabla^k\nabla_k\right]Y =: \frac{1}{1+p^2}\mathcal{D}^i_j Y. \quad (5.21)$$

We note that $Y^i_j$ satisfies the equations,

$$\nabla_i Y^i_j = -\frac{2}{3}\frac{4+p^2}{1+p^2}\nabla_j Y, \qquad \nabla^k\nabla_k Y^i_j = -(7+p^2)Y^i_j. \quad (5.22)$$

By an infinitesimal coordinate transformation $\xi \to \widetilde\xi = \xi + T(x)$, we have

$$\widetilde\varphi(x) = \varphi(x) - \dot\phi_B T(x), \quad \widetilde{\mathcal{R}}(x) = \mathcal{R}(x) - \frac{1}{\xi}T(x), \quad (5.23)$$

on the new hypersurface [18]. Hence setting $T = \varphi/\dot\phi_B$, we obtain

$$\widetilde{\mathcal{R}}(x) = -\frac{1}{\xi\dot\phi_B}\varphi(x), \quad (5.24)$$

as the curvature perturbation of the $\phi = const.$ hypersurface.

Now for $p^2 = -4$, one finds the trace of the curvature perturbation vanishes. Furthermore, from the first equation of (5.22), we see that $Y^i_j$ becomes transverse. Thus the $p^2 = -4$ scalar-type curvature perturbation happens to become transverse-traceless. This suggests that $\ell = 0$ and $1$ modes do not contribute to the curvature perturbation and the term violating the Lorentz invariance in Eq.(5.18) will disappear when the two-point function of the curvature perturbation is considered. In fact, this can be explicitly demonstrated by operating with $\mathcal{D}_{ij}$ on it. Consequently, we obtain

$$\langle\delta^{(3)}\widetilde R^i_j(x)\delta^{(3)}\widetilde R^k_l(x')\rangle_{p^2=-4} = \frac{1}{4\pi^2 N^{(2)}_{wall}\xi^3\xi'^3}\mathcal{D}^i_j\mathcal{D}'^k_l\left[\zeta\frac{\cosh 2\zeta}{\sinh\zeta} - \frac{11}{3}\cosh\zeta\right], \quad (5.25)$$

which is manifestly Lorentz covariant.

From the above discussion, it is anticipated that the quantum fluctuations of the scalar field inside the bubble will be better understood when we include degrees of freedom of the metric perturbation. This point will be disclosed in the future work.

Now let us examine if there exist discrete modes other than $p = 2i$. Since the mode (5.16) has no node, it is the eigenfunction for Eq. (2.24) with the lowest eigenvalue, i.e., the smallest $p^2$. Hence other possible discrete modes must have the eigenvalues in the range $-4 < p^2 < 0$ and have at least one or more nodes. Now in the thin-wall limit, it can be explicitly shown that such modes do not exist. Also in this limit, one finds the mode function $G_p$ with $p = 0$ has one node and diverges to minus infinity as $\xi_R \to 0$. Then as we vary the model parameters continuously, the value of the mode function $G_{p=0}$ at $\xi_R = 0$ should cross zero if there should appear a bound state mode with one node. This implies the divergence of $n_0$ for a certain set of the model parameters. However, as seen from Figs. 4 and 5, we have found no divergence of $n_0$. Hence in our model, we conclude that there exist no additional discrete modes other than those related to the wall fluctuations.

We will present a detailed analysis of these discrete modes in future publication.

## VI. CONCLUSION

In this paper, we have investigated the self-excitation of a scalar field in process of its decay from a false vacuum. For this purpose we have considered a model potential which is piece-wise quadratic, hence allows us analytical treatments. We have interpreted the resulting quantum state inside a nucleated bubble in the particle creation picture. Then we have found the following features of particle creation.

When the spacetime region of the bubble wall, which is defined as the region in which the mass-squared of the scalar field is negative, is confined in the Euclidean region $\mathcal{E}$, the number of created particles per each mode is exponentially suppressed in the thin-wall limit and at most of order unity unless the mass scale at the true vacuum is exponentially small compared with that at the top of the potential barrier. On the other hand when the wall region extends to the Lorentzian region $\mathcal{M}$, i.e., the mass-squared at the center of the bubble is still negative, the particle creation can be significantly enhanced. In this case, we have derived an approximate formula (4.22) for the particle spectrum as a function of the model parameters which determine the shape of the potential, where $\mathcal{F}_\phi$ is defined by Eq.(4.15).

In addition, we have also considered the effect of a set of discrete modes which describe the oscillation of the bubble wall. From a careful analysis of the Klein-Gordon norms, we have argued that these discrete modes do contribute to the quantum state inside the bubble, though they cannot be interpreted as usual particle modes. However, since the monopole and dipole components of these modes corresponds to the spacetime translation of the bubble center, we have argued that a consistent treatment of these modes requires the inclusion of gravitational degrees of freedom into the analysis simultaneously.

In view of the above considerations, the next step to be taken is to take gravity into account in the background bounce solution. A framework in this direction has been already done in Ref. [19]. Hence it should be fairly straightforward to extend the present analysis to the one on the curved spacetime background. Then the second step is to study the effect of gravitational degrees of freedom on the excitation of a tunneling scalar field. In this respect, we expect that a formalism developed in Ref. [20] for dealing with the fluctuations around the bounce solution with gravity should be useful. After this second step, the role of the discrete modes will be clearly and unambiguously understood.

Once these issues are settled, we will be able to talk about the quantum state of the scalar field inside the bubble with confidence. In connection with the one-bubble open inflation scenario, we will be able to discuss quantitatively the influence of the quantum fluctuations induced by tunneling on the primordial density perturbations and on the CMB anisotropies on large angular scales.

## ACKNOWLEDGMENTS

We would like to thank Prof. H. Sato for continuous encouragements. This work was supported by Monbusho Grant-in-Aid for Scientific Research Nos. 2841 and 5326 and 05640342.